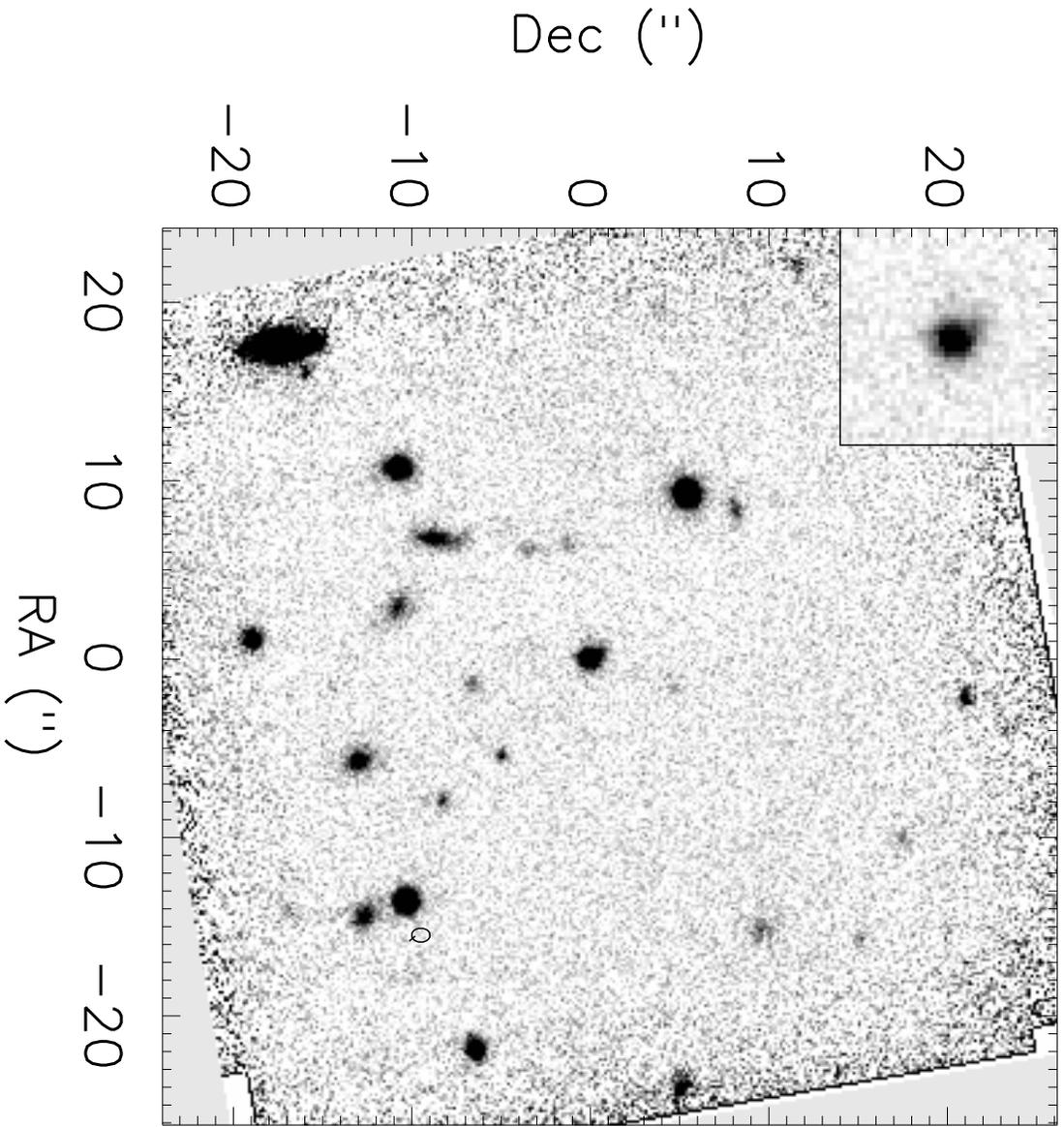

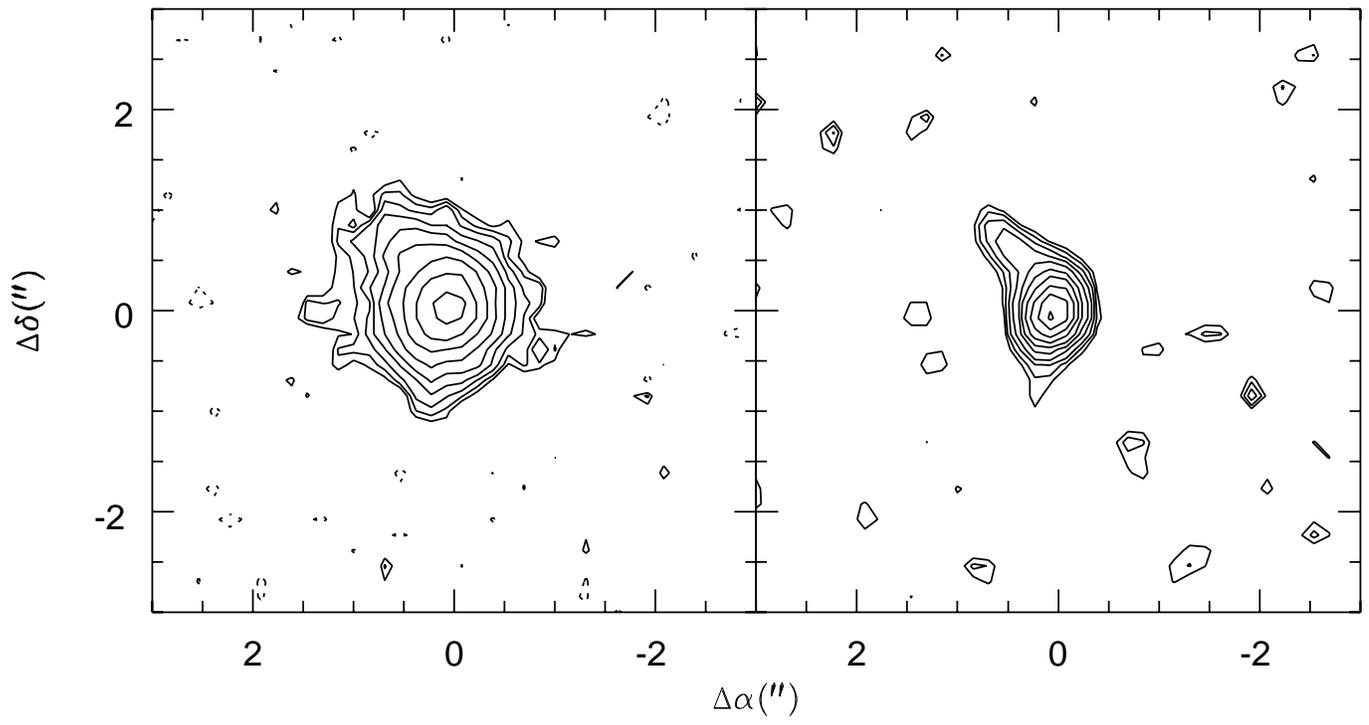

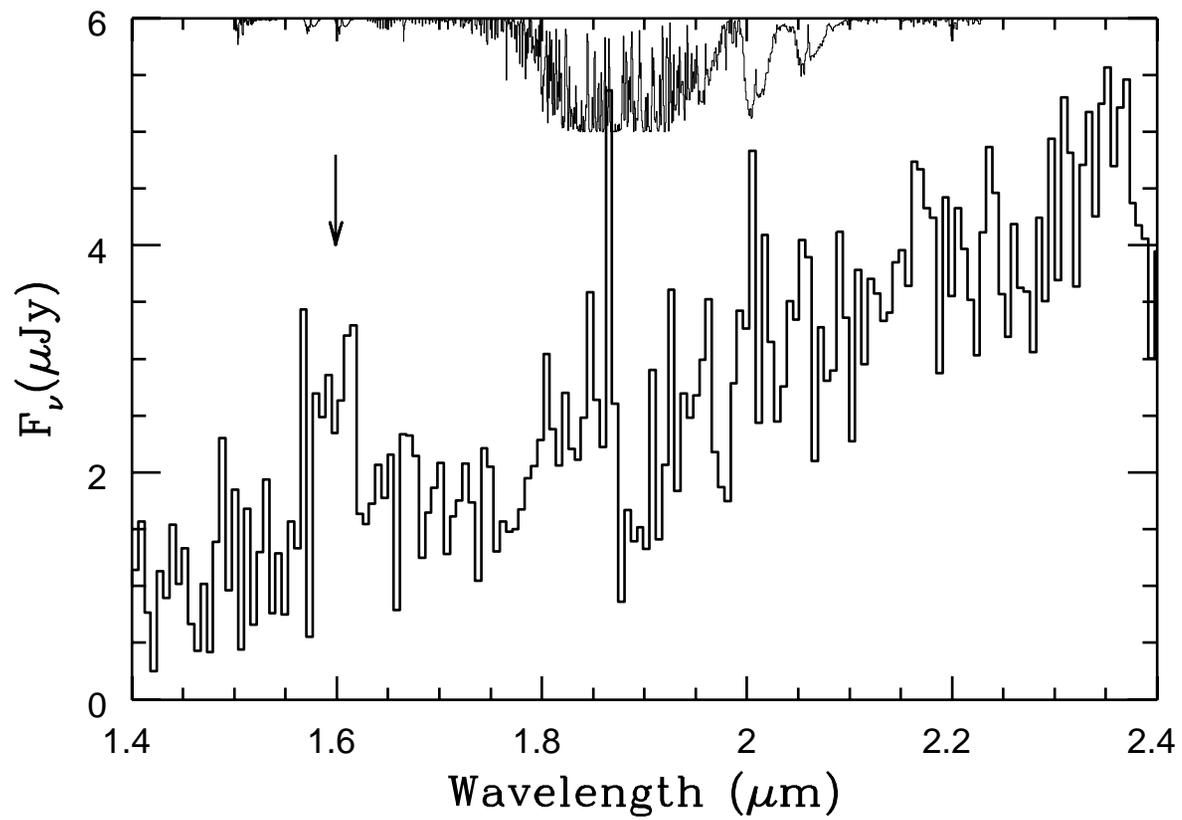

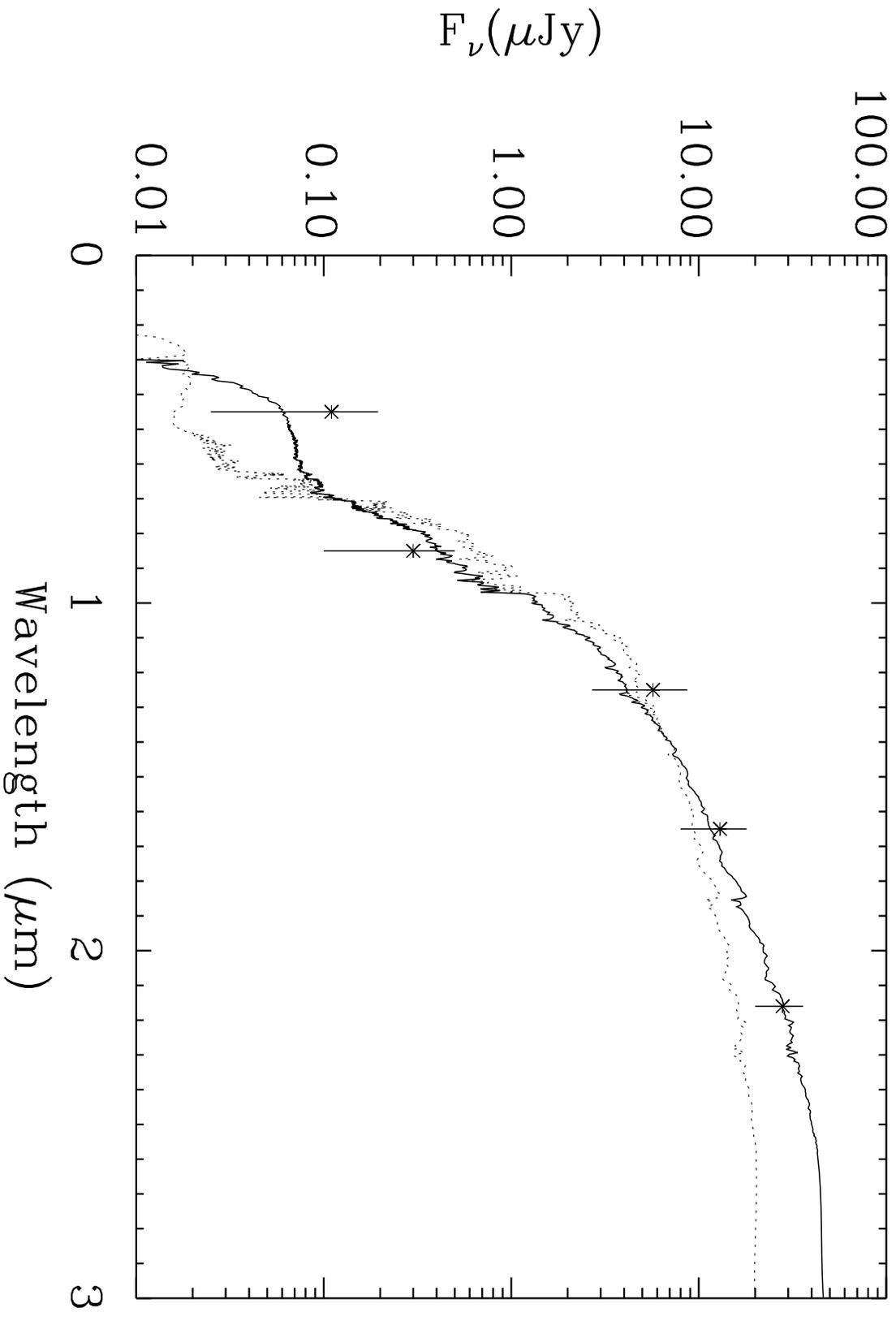

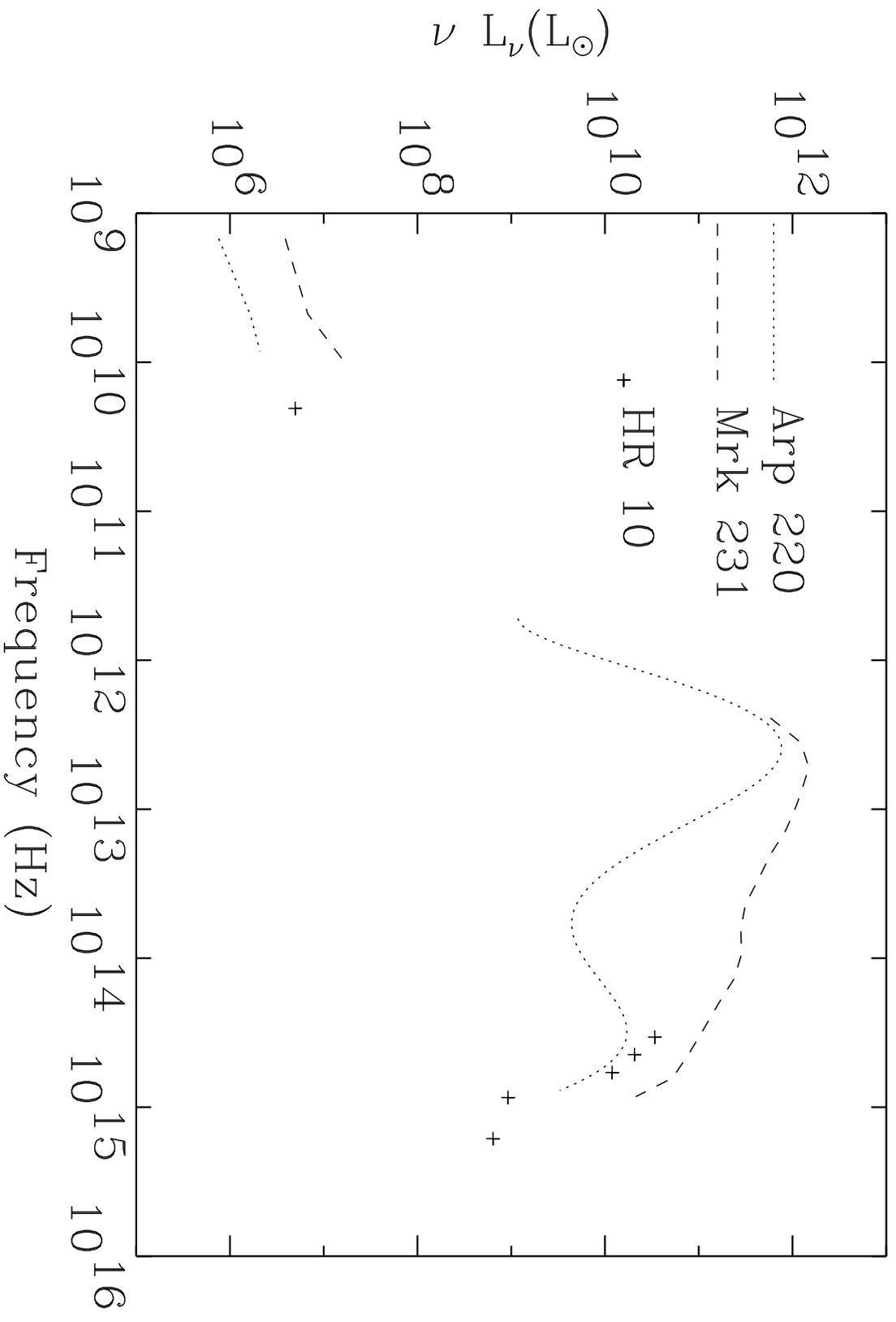

# THE SPECTROSCOPIC REDSHIFT OF AN EXTREMELY RED OBJECT AND THE NATURE OF THE VERY RED GALAXY POPULATION[1]

James R. Graham

Astronomy Dept., University of California at Berkeley, CA 94720

E-mail: jrg@graham.berkeley.edu

Arjun Dey

KPNO/NOAO[2], 950 N. Cherry Ave., P. O. Box 26732, Tucson, AZ 85726

E-mail: dey@noao.edu

## ABSTRACT

Infrared surveys have discovered a significant population of bright ($K \lesssim 19$) extremely red ($R - K \gtrsim 6$) objects. Little is known about the properties of these objects on account of their optical faintness ($R \gtrsim 24$). Here, we report deep infrared imaging and spectroscopy of one of the extremely red objects (EROs) discovered by Hu & Ridgway (1994) in the field of the $z = 3.79$ quasar PC 1643+4631A. The infrared images were obtained in $0\rlap{.}''5$ seeing, and show that the object (denoted HR 10) is not a dynamically relaxed elliptical galaxy dominated by an old stellar population as was previously suspected, but instead has an asymmetric morphology suggestive of either a disk or an interacting system. The infrared spectrum of HR 10 shows a single, possibly broad emission feature at 1.60 $\mu m$ which we identify as H$\alpha$+[N II] at $z = 1.44$. The luminosity and width of this emission line indicates either intense star formation ($\sim 20 \ h^{-2} \ M_\odot \ yr^{-1}$) or the presence of an active nucleus. Based on the rest frame UV-optical spectral energy distribution, the luminosity of HR 10 is estimated to be $3 - 8 \ L^*$. The colors of HR 10 are unusually red for a galaxy (at $z = 1.44$ the age of HR 10 is $2 - 8 \ Gyr$ depending on cosmology), and indicate that HR 10 is dusty. HR 10 is detected weakly at radio wavelengths; this is consistent with either the starburst or AGN hypothesis. If HR 10 is a typical representative of its class, EROs are numerous and represent a significant component of the luminous objects in the Universe at $z \approx 1.5$.

---

*Subject headings:* cosmology: early universe — galaxies: redshifts — galaxies: individual (HR 10) – galaxies: evolution — infrared: galaxies

## 1. Introduction

An intriguing new mystery emerging from several infrared imaging surveys is the discovery of bright ($K \lesssim 19$), extremely red ($R - K \gtrsim 6$), objects that appear quite common in the field. A significant population of red objects was first noted in the early $K$–band imaging surveys of Elston, Rieke & Rieke (1988, 1989). Since then, a surprising number of objects with $R - K \gtrsim 6$ have been identified in the vicinity of high redshift radio galaxies and quasars (McCarthy, Persson & West 1992; Eisenhardt & Dickinson 1992; Hu & Ridgway 1994; Graham et al. 1994; Dey, Spinrad & Dickinson 1994; Soifer et al. 1994). The EROs are extremely faint optically ($R \gtrsim 24$), and hence their redshifts and spectral properties (i.e., whether they are normal or active galaxies, old, or young and reddened) remain unknown.

These extremely red objects (EROs) are spatially extended, with angular sizes of about 0''5 and therefore are almost certainly galaxies. They were not selected by radio emission, but a few appear to be weak radio sources. EROs are also curiously abundant: their surface density is ≈0.01 per square arcmin, which makes this class of galaxies as common as quasars (Hu & Ridgway 1994, Cowie et al. 1994), and they therefore constitute a significant population in the Universe. There is also some marginal evidence that the EROs have a higher surface density (by at least an order of magnitude) in fields around AGN (Dey, Spinrad & Dickinson 1995).

The existence of a significant population of objects that is extremely red and bright is difficult to explain using the known properties of nearby galaxies. The extreme red colors of the EROs might be attributed either to an old stellar population, or to dust reddening, or to both. In the absence of spectroscopic evidence, the broad–band colors are ambiguous. If we assume that the colors are not affected by dust reddening, then the $R - K$ color and $K$ magnitude can be explained by redshifting present-day $L^*$ ellipticals to $z \approx 1 - 2$. If we include the possibility of dust, the colors and magnitudes may be explained by reddening of either a starburst population or an AGN. Finally, the extreme red color may also result if the $K$–band magnitude is contaminated by a strong emission line; the measured $(R - K)$ color is then unrepresentative of the true continuum SED. The lowest redshift possibility for this is $z \approx 2 - 2.5$, when H$\alpha$ is in the $K$–band. However, it is difficult to imagine why a large population of high redshift luminous objects would exist in this narrow redshift range, have strong Hα line emission and yet remain undetected in the Lyα searches. The only way to discriminate between these different possibilities is to obtain spectroscopic information for the EROs.

In this paper we present infrared observations of one of the reddest EROs, HR 10, discovered by Hu & Ridgway (1994) in their deep optical-infrared survey of the field of PC 1643+4631A (hereafter PC 1643). Our data consists of deep, high resolution infrared imaging and low resolution infrared spectroscopy of HR 10 obtained using the W. M. Keck telescope. We report a redshift for HR 10, the first for an ERO, and present a discussion of the nature of these mysterious objects based on our results for HR 10. Throughout we assume $H_0 = 100\ h\ km\ s^{-1}\ Mpc^{-1}$, and $q_0 = 0.5$. For this cosmology, the luminosity distance at $z = 1.44$ is $5264\ h^{-1}\ Mpc$ and $1''$ corresponds to $4.29\ h^{-1}kpc$. For $q_0 = 0.1$, the luminosity distance and scale are $6772\ h^{-1}\ Mpc$ and $5.51\ h^{-1}\ kpc/''$ respectively.

## 2. Observations and Reductions

### 2.1. IR Imaging

We observed PC 1643 on 1995 August 10 & 11 (UT) using the 10–m W. M. Keck telescope, Mauna Kea, Hawaii, with the facility near–IR camera (Matthews & Soifer 1994). The camera is equipped with a Santa Barbara Research Corporation $256 \times 256$ InSb array. The pixel scale is $0''.15$. We observed with the $K$ (2.0 – 2.4 $\mu m$) filter. An integration time of $10\ s$ per frame was used. Six frames were coadded before the telescope was offset by a few arcseconds using a non–redundant dither pattern. Each exposure was guided using an off–axis CCD camera.

Reduction of each coadded frame consisted of subtracting a sky frame constructed by averaging prior and subsequent frames. Objects in the sky frames were identified and excluded from the average. Sky–subtracted frames were then flat–fielded using an average of dark–subtracted twilight sky frames. We measured the positions of two to three stars common to all the reduced frames to determine the spatial registration and then shifted individual frames by integer pixel offsets to assemble a mosaic of the field. A total of thirty frames (1800 $s$) were obtained. The final mosaic is shown in Figure 1. A contour plot of HR 10 is shown in Figure 2a. The $K$–band mosaic achieves a $1\sigma$ limit of 23.52 magnitudes per square arcsecond. The images in the mosaic are seeing limited with $0''.5$ FWHM based on the measured size of the quasar, PC 1643.

## 2.2.   IR Spectroscopy

The Keck near–infrared camera is equipped with a focal plane slit mechanism and grisms, which can be used to obtain low resolution spectra. Spectra of HR 10 were obtained with a 120 $mm^{-1}$ grism blazed at 2.1 $\mu m$, a slit width of $0\rlap{.}''61$ yielding a resolution of 0.025 $\mu m$, and an $HK$ blocking filter that covers the wavelength range $1.4 - 2.5$ $\mu m$. Spectral observations consisted of nodding the telescope between discrete locations so that the HR 10 was located at five equally spaced positions separated by $6''$ along the slit. A single exposure of 90 $s$ was obtained at each telescope position. A total of 30 frames (2700 $s$) were obtained. To subtract the bright and time variable terrestrial OH $\Delta v = 2$ emission bands, which constitute the major source of background between 1.4 and 2.3 $\mu m$, a running average of the data was calculated and subtracted from each frame. Prior to subtraction, the average sky frame was scaled to minimize the variance of the sky subtracted frame. The sky subtracted frames were then shifted to a common spatial origin and averaged. The reduced spectrum in Figure 3 shows that this reduction procedure is very effective in canceling the airglow lines. The data in Figure 3 have also been flat fielded, and corrected for atmospheric absorption by using observations of the A0 star HD 160966.

## 3.   Results

## 3.1.   Imaging

The simplest explanation for the extreme red colors of objects such as HR 10 is that they are due to a quiescent stellar population, typical of current day elliptical galaxies seen at $z = 1 - 2$ (Graham et al. 1994, Hu & Ridgway 1994). The $K$–band image of the field of PC 1643 (Figures 1 and 2a) shows that HR 10 is not stellar, but is spatially extended with an intrinsic FWHM = $0\rlap{.}''65$ (2.8 $h^{-1}$ $kpc$ ). We have attempted to deconvolve the observed image using the IRAF/STSDAS implementation of the Lucy-Richardson image restoration algorithm (Lucy 1974). We used a kernel constructed by fitting elliptical isophotes to the image of the QSO PC1643; using this kernel, the QSO and fainter stars in the field devonvolve to point sources with FWHM $\approx 0\rlap{.}''28$. Figure 2b shows the Lucy deconvolved image of HR 10. HR 10 is not strongly nucleated and is distinctly asymmetric, with a pronounced extension in a direction $PA \approx 33° \pm 2°$, and a fainter extension to the south (in $PA \approx 168° \pm 2°$).

If the $K$ morphology is not due to the superposition of two or more objects, HR 10 is not an isolated, dynamically relaxed early-type galaxy. The asymmetric morphology at high surface brightness levels suggests that HR 10 may be either an interacting system or a disk galaxy with a prominent spiral (or tidal) arm. (We demonstrated that this latter explanation is viable by smoothing a $B$–band image of the nearby Sc galaxy M 99 to the same angular scale as HR 10; however, the deconvolution shown in Figure 2b is more indicative of an interacting system than a classical spiral arm.) In either case, it is probable that the spectral energy distribution (SED) of HR 10 cannot be interpreted without considering the effects of ongoing star formation and the associated dust extinction and reddening.

Alternatively, the asymmetric morphology of HR 10 may be due to the superposition of two (or more) distinct objects, with the isophotal extension in $PA \approx 20°$ being due to either a close companion galaxy or a random line-of-sight object. We note, however, that the extensions to both the NE and the south would require at least two companion (or line-of-sight) objects, and a more plausible explanation for the morphology may be that HR 10 is an interacting system. The present data are inadequate to address this issue, which requires HST optical and near-IR imaging and deep, high-spatial resolution spectroscopy. Hence, the $K$-band morphology, although suggestive, cannot be used alone to unambiguously rule out the possibility that HR 10 is a normal elliptical galaxy.

## 3.2. Spectroscopy

The grism spectrum of HR 10 (Figure 3) shows the flux rising steeply towards the red ($F_\nu \propto \nu^{-3.1\pm0.1}$). An emission feature is detected (at the $\approx 7\sigma$ level) at 1.60 $\mu m$ and is marginally resolved with a FWHM = $7000 \pm 3000$ $km\,s^{-1}$. If the emission feature is resolved, it may be intrinsically broad, or the width may due to a blend of distinct emission features. The equivalent width of the feature is $W_\lambda = 600 \pm 100$ Å (in the observed frame), and accounts for about 20% of the $H$–band flux. The emission line does not significantly perturb the IR photometry, and is therefore not the cause of the extremely red broad–band colors of HR 10.

H$\alpha$ +[N II] is the only plausible line identification. Other candidates, such as [O II]$\lambda3727$ or H$\beta$+[O III], can be ruled out due to the lack of emission features between 2.0 $\mu m$ and 2.4 $\mu m$. For example, if the emission line is [OIII]$\lambda5007,4959$ at $z \approx 2.2$, we would expect to see H$\alpha$+[NII] at $\lambda \approx 2.1$ $\mu m$. There is no emission line detected at this wavelength, and the upper limit requires an emission line ratio of [OIII]/H$\alpha \gtrsim 6$, which would be unprecedented for any normal star forming galaxy or AGN (Kennicutt 1992). Similarly, the emission line is unlikely to be [OII]$\lambda3727$ at $z \approx 3.3$, since this would imply an emission line ratio of [OII]/[OIII] $\gtrsim 6$; moreover, at this high redshift, HR 10 would be an extremely luminous ($> 50$ $L^*$) galaxy.

If we adopt the Hα + [N II] identification for the 1.60 $\mu m$ feature, then the redshift of HR 10 is 1.44 ± 0.01. The width of the observed emission feature (FWHM = 7000 ± 3000 $km\ s^{-1}$) cannot be due purely to blending of Hα and [N II] alone, because the separation of the two [N II] lines is only ≈ 1600 $km\ s^{-1}$. The line width, if real, may be then indicative of the presence of an active nucleus. However, given the low signal-to-noise ratio and low resolution of the present data, the measured line width is uncertain, and this conclusion is premature.

Confirmation of this redshift must await higher resolution spectroscopy (to resolve the Hα and [NII] lines), or the detection of either Hβ+[O III] at ≈ 1.22 $\mu m$ in the $J$–band atmospheric window or [O II] at ≈ 9100 Å. In the $HK$ wavelength region covered by our present spectrum, we may expect to detect [SIII]λλ9069, 9531. These lines are detected in most starbust, LINER and Seyfert galaxies: the ratio of the [S III] blend to Hα+[N II] lies in the range 0.03-0.5 with a mean value of 0.18 (Diaz et al. 1985, Kirhakos & Phillips 1989, Osterbrock, Tran & Veilleux 1992). If HR 10 is at $z = 1.44$, then our data yields [SIII]λλ9069, 9531/Hα + [N II] < 0.5 (3σ limit). If HR 10 is dusty and the spectrum is affected by reddening, the upper limit decreases. However, the present limit does not constrain the redshift, and is therefore consistent with the identification of the 1.60 $\mu m$ feature as Hα + [N II].

### 3.3.  Radio Properties

The field of PC 1643 has been observed with the VLA at 1.4 GHz by Schmidt et al. (1995) and at 8.44 GHz by Frayer (1996). These maps were kindly made available to us by Drs. van Gorkom and Frayer. HR 10 is detected at the ≈ 3σ level at 8.44 GHz, but not detected at 1.4 GHz. (HR 14, the other ERO in the field, is also marginally detected – about 2σ – in the 8.44 GHz map, but not in the 1.4 GHz map.) The measured flux density of HR 10 at 8.44 GHz is 35 ± 11 $\mu Jy$ (D. Frayer, pers. comm.), which corresponds to a rest frame radio luminosity of $L_{20.6GHz} \approx 4.8 \times 10^{22}\ h^{-2}\ W\ Hz^{-1}$. The noise in the 1.4 GHz map is about 100 $\mu Jy/beam$; thus a 3σ upper limit on the rest frame radio luminosity is $L_{3.7GHz} < 4 \times 10^{23}\ h^{-2}\ W\ Hz^{-1}$. Hence, the 1.4 GHz upper limit only loosely constrains the radio spectral index to $\alpha < 1.2$ (where $S_\nu \propto \nu^{-\alpha}$), and implies rest frame luminosities for HR 10 of $L_{1.4GHz} \approx 0.5 - 11 \times 10^{23}\ h^{-2}\ W\ Hz^{-1}$ and $L_{8.44GHz} \approx 0.5 - 1.4 \times 10^{23}\ h^{-2}\ W\ Hz^{-1}$ for $0 < \alpha < 1.2$. HR 10 is therefore a radio quiet object (i.e., since radio loud objects typically have $L_{1.4GHz} > 2.5 \times 10^{24}\ h^{-2}\ W\ Hz^{-1}$). We note, however, that the derived radio power does not place very stringent restrictions on the nature of HR 10: the monochromatic radio powers for Seyfert galaxies are $L_{1.4GHz} = 3.5 \times 10^{19} - 4.3 \times 10^{24}\ h^{-2}\ W\ Hz^{-1}$ (Wilson & Heckman 1985), and a typical ultraluminous $IRAS$ galaxy is characterized by a radio luminosities of $L_{1.4GHz} \sim 6 \times 10^{22} h^{-2}$ W $Hz^{-1}$ and $L_{8.44GHz} \sim 2 \times 10^{22} h^{-2}$ W $Hz^{-1}$ (Condon et al. 1991). In comparison, Arp 220 has a rest frame 8.44 GHz luminosity of $6 \times 10^{22} h^{-2}$ W $Hz^{-1}$ (Condon et al. 1991).

## 4. Discussion

### 4.1. What is HR 10?

Hu and Ridgway (1994) argued that the red color of HR 10 could be accounted for by an $\approx 10L^*$ elliptical galaxy at $z \approx 2.4$. However, the asymmetric $K$–band morphology of HR 10 (Figure 2) and the detection of a high equivalent width ($W_\lambda \approx 600$ Å) emission line at 1.6 $\mu m$ are both inconsistent with HR 10 being a quiescent elliptical galaxy. If our identification of the emission line as H$\alpha$+[NII] is correct, and the redshift is 1.44, then an unreddened elliptical galaxy provides a very poor match to the observed SED (Fig. 4; dotted line). The most significant discrepancy is due to the extreme redness of the observed SED between $I$ and $K$. The opposite is true between $B$ and $I$: HR 10 is too blue to match an elliptical galaxy. We find a lower limit to the luminosity of HR 10 of $L = 3.3$ $L^*$ by using this elliptical galaxy SED to calculate a k-correction to apply to our IR photometry.

If we redden the elliptical SED ($A_V \approx 1.1$), then it is possible to get a good fit to the $I$ through $K$ data, but then the fit to $B$ is much worse. HR 10 is only very marginally detected in the $B$ band data of Hu and Ridgway (1994), and its $B$ flux is fairly uncertain. Nevertheless, taking the $B$ magnitude at face value, it appears that the ingredients necessary to reproduce the entire SED of HR 10 are extinction and recent star formation or AGN activity. As noted in §3.1, HR 10's asymmetric morphology is consistent with that of a face-on spiral galaxy or an interacting system. The presence of an emission line also suggests that its SED may be fit better by a galaxy with star formation or AGN activity. For example, the observed broad–band fluxes can be well fit by the SED of a model Sb galaxy plus screen extinction (Bruzual and Charlot 1993). The solid line in Fig. 4 shows the best fit Sb SED which is found for $A_V = 1.8$. Integrating the best fit dereddened Sb SED yields $L = 8 \times 10^{10}$ $h^{-2}$ $L_\odot$, or $8L^*$ assuming $L^* = 1.0 \times 10^{10}$ $h^{-2}$ $L_\odot$ (Kirshner et al. 1983).

Using the SED's of well–observed nearby galaxies has the advantage of being a reliable empirical comparison. However, the SED's of present day galaxies do not make ideal templates because they contain stars that are very much older than HR 10. Therefore we have also compared the SED of HR 10 with the theoretical SEDs of instantaneous starbursts from the models of Bruzual and Charlot (1993). We used a Salpeter initial mass function (IMF) with a mass range of $0.1 - 125\ M_\odot$. This comparison provides no useful limit on the age of HR 10, but shows once more that that the SED cannot be fit without extinction. Our results may be parametrized by the relation $A_V = 2.1 - 0.8 log_{10}(t_9) \pm 0.3$ for $0.1 < t_9 < 10$, where $t_9$ is the age of the starburst in $Gyr$. Constraints on the age of HR 10 from the SED fit are poor, but a crude upper limit to its age may be derived by assuming formation at $z = \infty$; the age then lies in the range from $1.7\ Gyr$ ($h = 1$, $q_0 = 1/2$) to $8.0\ Gyr$ ($h = 0.5$, $q_0 = 0$) and hence $A_V = 1.8 - 2.5$, with a statistical uncertainty of 0.3.

It is therefore inescapable that HR 10 is dusty. The color of a stellar population depends on several factors including age, IMF, metallicity and star formation history. However, only extreme measures, such as truncating the upper mass limit of the IMF at about $0.4\ M_\odot$, can produce an intrinsic SED as red as HR 10. An alternate star formation history cannot make an SED red enough to explain HR 10 either. This is because a composite population (i.e., one with episodes of star formation) contains, on average, younger, bluer stars than those formed in an instantaneous burst. Hence, for a given age, IMF, and metallicity, the color of an instantaneous burst provides an upper limit on the redness of a stellar population. Finally, we note that the SED shown in figure 4 is determined from spatially averaged fluxes and therefore clearly comes from a composite population: a comparison of the Hu & Ridgway (1994) $B$–band image and the Keck $K$–band image shows that the optical and IR emission are not cospatial. The nuclear regions of HR 10 are therefore even redder and dustier than suggested by the integrated photometry.

The strong H$\alpha$+[N II] line indicates either enhanced star formation or the presence of an active nucleus. The rest frame H$\alpha$+[N II] equivalent width is $250 \pm 40$ Å, which is much larger than the mean H$\alpha$+[N II] equivalent width of 29 Å observed in Sc and SBc galaxies (Kennicutt & Kent 1983). The large equivalent width found in HR 10 is usually only associated with galaxies that are classified as Im or starbursts (Kennicutt 1992) or galaxies with active nuclei (Osterbrock 1979). The total line luminosity, $L_{H\alpha+[N\ II]} = 3 \times 10^{42}\ h^{-2}\ erg\ s^{-1}$, is an order of magnitude larger than the H$\alpha$ luminosity of a typical giant Sc ($2.5 \times 10^{41}\ h^{-2}\ erg\ s^{-1}$, Kennicutt 1992) and compares only with the most extreme examples. If HR 10 harbors a starburst and the line emission is from H II regions ionized by hot young stars, then the observed line luminosity corresponds to a star formation rate for massive stars ($M > 10\ M_\odot$) of $3\ h^{-2}\ M_\odot\ yr^{-1}$ and a total rate of $20\ h^{-2}\ M_\odot\ yr^{-1}$ (Kennicutt 1983).

HR 10's line luminosity is also in accord with that of a Seyfert II ($L_{H\alpha} = 10^{40-42} erg\ s^{-1}$) or a Seyfert I/quasar ($L_{H\alpha} = 10^{42-46} erg\ s^{-1}$). The strength of HR 10's line emission is therefore consistent with either starburst or nuclear activity, but without additional line diagnostics it is impossible to decide which. There is marginal evidence that the intrinsic width of Hα+[N II] is large (FWHM $\approx 7000 \pm 3000 \; km \; s^{-1}$) — this width is more consistent with an AGN rather than a starburst origin for the strong emission line. However, to conclude that HR 10 has broad lines characteristic of a Seyfert I or a quasar would be premature given the low resolution and signal to noise ratio of the present data. Thus taken as a whole, the morphology, SED, and spectrum suggest that HR 10 is not an isolated, dynamically relaxed elliptical galaxy, but instead either a luminous, dusty starburst galaxy or a dusty AGN.

Whether it is an AGN or a starburst galaxy, HR 10 is a dusty system. Hence, a substantial fraction of its luminosity may emerge at far–infrared (FIR) wavelengths. In the local universe such galaxies are represented in the sample of *IRAS* bright galaxies (Soifer et al. 1987). Figure 5 compares the rest frame SED of HR 10 compared with Arp 220 ($L_{FIR} = 0.9 \times 10^{12} \; h^{-2} \; L_\odot$) and Mrk 231 ($L_{FIR} = 1.9 \times 10^{12} \; h^{-2} \; L_\odot$), two archetypical low redshift examples of the most luminous objects detected by *IRAS*, the so-called ultraluminous *IRAS* galaxies ($L_{FIR} > 0.6 \times 10^{12} h^{-2} \; L_\odot$) (Sanders et al. 1988). These ultraluminous galaxies are systems with intense star formation or AGN activity (or both) fueled by abundant molecular gas which has been funneled into the nuclear regions by interactions or mergers. It is intriguing that the rest frame optical colors of HR 10 and Arp 220 are very similar (Figure 5), although HR 10 is more luminous. If the spectral similarity between HR 10 and Arp 220 extends to FIR wavelengths then HR 10 may have a luminosity of about $2.5 \times 10^{12} \; h^{-2} \; L_\odot$ which greatly exceeds that suggested by the rest frame optical data alone. On the basis of this comparison we would expect that the SED of HR 10 would peak at 200 $\mu m$ at $F_\nu \approx 100 \; mJy$. This is below the detectability limit for *IRAS*, but readily detectable by the ISOPHT instrument on ISO (ISO Science Operations Team, 1991).

## 4.2. The Space Density of the EROs

We have presented the very first measurement of the redshift of an ERO. In this speculative subsection, we investigate the consequences of the hypothesis that all the known EROs are intrinsically similar to HR 10. Most of the known EROs have similar apparent magnitudes ($K \approx 18.5$) and colors ($R - K \gtrsim 6$), and comparable angular sizes ($\theta \approx 0\rlap{.}''5$), which together suggest that these objects may be at similar redshifts. The obvious caveat here is that the known EROs may well constitute a heterogeneous sample: their observed properties may be alike not because of intrinsic similarity, but only because our selection criteria have chosen the brightest and reddest of members of unrelated populations. In particular, we have determined the redshift of the reddest ERO known; it is entirely plausible that the colors of the reddest EROs are determined by dust reddening, whereas the colors of the less extreme EROs ($R - K \approx 6$) are due to old stellar populations. Nevertheless, pleading ignorance, in this subsection we adopt the simplifying assumption that the EROs form a homogeneous population. Some weak support for this assumption is provided by the idea that most of the EROs are unlikely to be at a very high redshifts ($z > 3$), since this would require that these objects be exotic and very luminous galaxies. It is also unlikely that most of the EROs are at low redshift, since no population with the properties of EROs is yet known to exist locally.

If the EROs do indeed form a homogeneous population, and the redshift of HR 10 is typical of this class, then we can estimate the space density of these objects. We consider six objects selected from four fields (MG 1019+0535; Dey, Spinrad, & Dickinson 1995, B2 0902+34 ; Eisenhardt & Dickinson 1992, 4C 41.17; Graham et al. 1994, and PC 1643; Hu & Ridgway 1994). The space density was calculated based on considering $V_{max}$, the maximum volume to which each object could have been detected (Schmidt 1968). Application of this method requires a knowledge of the SED to calculate k-corrections. Here we have made the simplifying assumption that the spectrum is flat in $F_\nu$, even though the spectrum actually rises steeply towards the red, in order to make the most conservative estimate of the space density. We find that the space density of EROs is $\rho = 1.7 \pm 0.7 \times 10^{-3} h^3 \ Mpc^{-3} \ mag^{-1}$. Under the assumptions made here (that HR 10 is typical of the EROs and there is no FIR contribution), the EROs have luminosities of $L \approx 3 - 8L^*$. Their space density is therefore not inconsistent with that of normal local galaxies of similar luminosity (Kirshner et al. 1983). This space density is also roughly consistent with that recently derived for luminous galaxies from the Canada-France Redshift Survey (CFRS; Lilly et al. 1995), and is also consistent with the CFRS result that the luminosity function of red galaxies shows very little evolution out to redshifts of $z \approx 1$. If HR 10 is a typical representative of this class, then the EROs constitute a significant fraction of the objects in this luminosity bin, provided there has been no strong evolution between $z = 1.4$ and the current epoch.

We have speculated that the SED of HR 10 might include a significant FIR component. If this were the case, then the EROs are more abundant than the *IRAS* ultraluminous galaxies and quasars by several orders of magnitude. If so, the EROs may be truly exotic objects undergoing a brief period of extreme activity. This is such a bizarre conclusion that it would seem to exclude this possibility.

The surface density of these objects is measured to be higher (by at least an order of magnitude and perhaps by as much as a factor of 100) in fields around high redshift radio galaxies and quasars than in the general field (Dey, Spinrad & Dickinson 1995). However, there is only a small number of EROs known, and their surface density is therefore highly uncertain. However, if their space density in the fields around radio galaxies and quasars *is different* from that in the general field, then the red galaxies be may clustered around luminous AGN. This is unlikely given the relatively bright apparent magnitudes of the red objects – if they are objects at redshifts $\approx$ 2–3 they would be $15 - 20$ times more luminous than $L^*$ ellipticals at the same redshift. If instead they are members of foreground clusters, their increased density in AGN fields may imply that our samples of luminous AGN are biased by gravitational lensing.

## 5. Conclusions

We have obtained deep $K$–band imaging and low resolution infrared spectroscopic observations of HR 10, one of the extremely red objects ($I - K \approx 6.5$) discovered by Hu & Ridgway (1994). The asymmetric $K$–band morphology, SED and strong line emission of HR 10 all suggest that HR 10 is not a quiescent elliptical galaxy, but instead a luminous interacting galaxy at $z = 1.44$. The line emission and possible nonthermal radio emission are indicative of either an ongoing luminous starburst or an active nucleus. If the other extremely red objects discovered in IR imaging surveys are similar to HR 10, then the volume space density of these objects may be larger than that of the ultraluminous *IRAS* galaxies and quasars.

We thank W. Harrison and B. Schaeffer for invaluable help during our Keck run and to I. de Pater for the telescope time to make these observations. We are grateful to M. C. Liu for his help with the IR reductions. We are grateful to E. Hu for providing us with her images of the PC 1643 field, and for useful discussions. We thank J. van Gorkom and D. Frayer for providing us with their radio observations of PC 1643 and for useful discussions, and to T. Lauer for comments on an early version of the manuscript. We thank N. Trentham for supplying us with the tabulated SED for Arp 220 and Mrk 231. We are grateful to H. Spinrad for his encouragement of this project, and to the referee P. J. McCarthy for constructive comments. This research made use of the NASA/IPAC Extragalactic Database which is operated by JPL, Caltech, under contract with NASA. The W. M. Keck Observatory is a scientific partnership between the University of California and the California Institute of Technology, made possible by a generous gift of the W. M. Keck Foundation. JRG's research is supported in part by the Packard Foundation.

Table 1.  Photometry of HR 10

| Observed Band | Rest Wavelength | Magnitude | $F_\nu$ ($\mu$Jy) | Reference |
|---|---|---|---|---|
| $B$ | 1800Å | $26.11^{+0.49}_{-0.34}$ | $0.16 \pm 0.07$ | 1 |
| $I$ | 3480Å | $24.44^{+0.33}_{-0.25}$ | $0.41 \pm 0.13$ | 1 |
| $J$ | 4920Å | $21.00^{+0.35}_{-0.26}$ | $6.4 \pm 2.1$ | 1 |
| $H$ | 6560Å | $19.62^{+0.26}_{-0.21}$ | $14.8 \pm 3.6$ | 1,2 |
| $K$ | 9010Å | $18.42^{+0.03}_{-0.02}$ | $27.7 \pm 0.6$ | 2 |
| 1.4 GHz | 8.6cm | $\cdots$ | $35 \pm 11$ | 3 |
| 8.4 GHz | 1.5cm | $\cdots$ | $< 300$ | 3 |

Note. — The rest wavelengths are estimated using $z = 1.44$. All magnitudes are measured in a circular aperture of diameter $3''$. Errors quoted are $1\sigma$. References: [1] remeasured from the Hu & Ridgway (1992) data; [2] this paper; [3] Frayer 1996.

Table 2.  Spectral Data

| $\lambda_{obs}$ | Flux (erg s$^{-1}$ cm$^{-2}$) | FWHM ($km\,s^{-1}$) | $W^{obs}_\lambda$ (Å) |
|---|---|---|---|
| $1.60 \pm 0.01 \mu m$ | $8.9 \pm 0.2 \times 10^{-9}$ | $7000 \pm 3000$ | $600 \pm 100$ |

## Figure Captions

Fig. 1.— Keck $K$–band image of the PC 1643+4631A field. This mosaic has an effective seeing FWHM of $0\rlap{.}''5$ and is centered on the red galaxy HR 10. The coordinate offsets are relative to HR 10. The inset in the top left corner shows a detail of HR 10, $6''$ on a side. HR 10 is resolved, with an intrinsic FWHM of $0\rlap{.}''7$, and has an asymmetric structure. The QSO PC 1643+4631A is located at $(-14'', -11'')$ and labeled Q. Note the faint companion (a candidate for the damped Ly$\alpha$ absorber?) and the low surface brightness bridge that runs from the quasar to the south. At $z = 1.44$, $1''$ corresponds to $4.29$ $h^{-1}$ $kpc$.

Fig. 2.— (a) Contour plot of the Keck $K$–band image of HR 10 showing the asymmetric structure to the NE of the galaxy. The solid contours are drawn at levels $(2,3,4,5,6,7,8,9,10) \times \sigma_{sky}$, where $\sigma_{sky} = 22.28$ $mag/arcsec^2$. The dotted line represents the $-2\sigma_{sky}$ contour level. (b) Lucy deconvolution of the $K$-band image using a psf constructed from the QSO PC1643+4631A. The contour levels are drawn as in (a), but with $\sigma_{lucy}$ equivalent to $21.38$ $mag/arcsec^2$. The deconvolved image has a resolution FWHM of $\approx 0\rlap{.}''28$.

Fig. 3.— Keck IR spectrum of HR 10 showing a broad emission feature at $1.6\mu$m (indicated by the arrow). The most likely identification for the emission feature is H$\alpha$+[NII] at $z \approx 1.44$. The atmospheric transmission spectrum extracted from the National Solar Observatory IR Solar Atlas (Livingston & Wallace 1991) is shown as the light line at the top of the plot. The atmospheric curve has been shifted such that the transmission is unity at the top of the plot, and zero at the $5\mu Jy$ value of the ordinate.

Fig. 4.— Broad-band photometry of of HR 10 compared with the best fit model Sb galaxy SED (solid line) at $z = 1.44$ ($A_V = 1.8$). The dotted line is the best fit unreddened elliptical galaxy SED at $z = 1.44$.

Fig. 5.— Comparison of the rest frame SED of HR 10, Arp 220 and Mrk 231. The SEDs for Arp 220 and Mrk 231 are from N. Trentham and use data from Sanders et al. (1988), Condon & Broderick (1991) and David et al. (1992).